\documentclass[%
12pt,T
 oneocolumn,
 nofootinbib,
 amsmath,amssymb,
 aps,
floatfix,
unsortedaddress
]{revtex4-2}

\usepackage{graphicx,color}
\usepackage{subfig}
\usepackage{dcolumn}% Align table columns on decimal point
\usepackage{bm}% bold math
\usepackage{graphicx,color}
\usepackage{float}
\usepackage{amsmath}
\usepackage{mathtools}
\usepackage{cancel}
\usepackage[pdfpagelabels, pdfencoding=auto, psdextra]{hyperref}
\hypersetup{%
 pdfsubject=Paper,
 pdfkeywords={nuclear physics} {Pionless EFT} {Two-Nucleon} {Lattice} {L\"usher formula},
 unicode = true,
 breaklinks = true,
 colorlinks = true,
 linkcolor = blue,
 citecolor = blue,
 menucolor = blue,
 citecolor = blue,
 urlcolor = blue
}

\begin{document}

\title{Coupled-channel $\Lambda_{c}K^{+}-pD_{s}$ Interaction in Flavor $ \textrm{SU}\left(3\right) $ Limit of Lattice QCD }

\author{Faisal Etminan}
 \email{fetminan@birjand.ac.ir}
\affiliation{ Department of Physics, Faculty of Sciences, University of Birjand, Birjand 97175-615, Iran
}%
\affiliation{ Interdisciplinary Theoretical and Mathematical Sciences Program (iTHEMS), RIKEN, Wako 351-0198, Japan}

\author{Kenji Sasaki }
 %\email{}
\affiliation{ Center for Gravitational Physics, Yukawa Institute for Theoretical Physics, Kyoto University, Kyoto 606-8502, Japan}

%\author{Takaya Miyamoto }
%\email{}
%\affiliation{ Center for Gravitational Physics, Yukawa Institute for Theoretical Physics, Kyoto University, Kyoto 606-8502, Japan}

%\author{Sinya Aoki }
%\email{}
%\affiliation{Center for Gravitational Physics, Yukawa Institute for Theoretical Physics, Kyoto University, Kyoto 606-8502, Japan}

%\author{Takumi Doi }
%\email{}
%\affiliation{Interdisciplinary Theoretical and Mathematical Sciences Program (iTHEMS), RIKEN, Wako 351-0198, Japan}

%\author{Tetsuo Hatsuda }
%\email{}
%\affiliation{Interdisciplinary Theoretical and Mathematical Sciences Program (iTHEMS), RIKEN, Wako 351-0198, Japan}

%\author{Etsuko Itou }
%\email{}
%\affiliation{Center for Gravitational Physics, Yukawa Institute for Theoretical Physics, Kyoto University, Kyoto 606-8502, Japan}
%\affiliation{Interdisciplinary Theoretical and Mathematical Sciences Program (iTHEMS), RIKEN, Wako 351-0198, Japan}

%\author{Yoichi Ikeda }
%\email{}
%\affiliation{Research Center for Nuclear Physics (RCNP), Osaka Univ., Osaka 567-0047, Japan}

\author{Takashi Inoue }
%\email{}
\affiliation{College of Bioresource Science, Nihon Univ., Kanagawa 252-0880, Japan}

%\author{Yan Lyu }
%\email{}
%\affiliation{State Key Laboratory of Nuclear Physics and Technology, 	School of Physics, Peking University, Beijing 100871, China}
%\affiliation{Interdisciplinary Theoretical and Mathematical Sciences Program (iTHEMS), RIKEN, Wako 351-0198, Japan}

%\author{Noriyoshi Ishii }
%\email{}
%\affiliation{Research Center for Nuclear Physics (RCNP), Osaka Univ., Osaka 567-0047, Japan}

%\author{Hidekatsu Nemura }
%\email{}
%\affiliation{Research Center for Nuclear Physics (RCNP), Osaka Univ., Osaka 567-0047, Japan}

%\author{(HAL QCD Collaboration) }
%\email{}
%\affiliation{}

\date{\today}% 

\begin{abstract}
We study $S$-wave interactions in the $I\left(J^{p}\right)=1/2\left(1/2^{-}\right)$
$\Lambda_{c}K^{+}-pD_{s}$ system on the basis of the coupled-channel
HAL QCD method. The potentials which are faithful to QCD S-matrix
below the $ pD^{*} $ threshold are extracted from Nambu-Bethe-Salpeter
wave functions on the lattice in flavor $ \textrm{SU}\left(3\right) $ limit. For the simulation, we employ $ 3 $-flavor full QCD 
gauge configurations on a $\left(1.93 \:\textrm{fm} \right)^{3}$ volume at $m_{\pi}\simeq 872$
MeV. 
%\textcolor{red}{For the charm quark, the relativistic heavy quark action is employed to treat its dynamics on the lattice}.
 We present our results of the
S-wave coupled-channel potentials for the $\Lambda_{c}K^{+}-pD_{s}$
system in the $1/2\left(1/2^{-}\right)$ state as well as scattering
observables obtained from the extracted potential matrix. We observe
that the coupling between $\Lambda_{c}K^{+}$ and $pD_{s}$ channels
is weak. The phase shifts and scattering length obtained from the
extracted potential matrix show that the $\Lambda_{c}K^{+}$ interaction
is attractive at low energy and stronger than the $pD_{s}$ interaction
though no bound state at $m_{\pi}\geq872$ MeV.
\end{abstract}

%\begin{keyword}
%Lattice QCD \sep  coupled-channel HAL QCD \sep potential \sep phase shifts
%\end{keyword}

\maketitle

%%%%%%%%%%%%%%%%%%
\section{Introduction}
The study of interactions between a charmed meson and a nucleon (N) is
an active field to investigate the property of the charmed hadronic matters.  
The interaction between a baryon and a meson due to the residual  strong force (nuclear force) can form bound, resonance or molecular states.  

Currently, the observation of new  XYZ states~\cite{BRAMBILLA20201} with hidden charm and/or beauty~\cite{RevModPhys.90.015004}, also the charmed tetraquarks~\cite{lhcb2022study,IkedaPRL2016,YanPRL2023}, and further
pentaquark states~\cite{PhysRevLett.122.222001}, make the studies of systems contain
charm hadrons like $ D $ and $ D^{*} $ more desirable. Very recently, for the first
time, the ALICE Collaboration measures the interaction between nucleons
and charm hadrons through the femtoscopic analysis, i.e., the two-particle momentum correlations of
$\textrm{pD}^{-}$ pair in high-multiplicity $\textrm{pp}$ collisions~\cite{PhysRevD.106.052010}.
%Furthermore, now the heavy flavor hadrons, i.e., charm and bottom, are investigated for future experiments at GSI-FAIR, J-PARC, and so on ~\cite{frimanLNP2011,HOSAKA201788}.

Furthermore, it has been known that the presence of heavy hadrons (those contain charm and bottom quarks) as impurities in a nuclear or quark matter leads to occurrence of the Kondo effect as well (in the case of quark matter it is known as QCD Kondo effect)~\cite{Yasuiprc2013,PhysRevD.92.065003,PhysRevD.105.074028}, this effect was originally observed in metal including impurities in the context of condensed-matter physics~\cite{Kondoptp1964,hewson_1993}. The Kondo effect changes the thermodynamic and transport properties of the quark/nuclear matter by converting
perturbative interaction at high energy scale to non-perturbative one at low energy scale in medium.
% the weak interaction at high energy scale to the strong interaction at low energy scale in medium.
 To be more specific, the Kondo effect due to existence of $  \Sigma_{c}\left(\Sigma_{c}^{*}\right) $ baryons~\cite{Yasuiprc2016,Yasuiprc2019} and $ \bar{D}_{s}\left(\bar{D}\right) $ mesons~\cite{YasuiPRC2017,YamaguchiPRD2022} as heavy impurities in nuclear matter are calculated while the spin and/or isospin exchange provides the non- Abelian interaction correspond to the Kondo effect. It is also demanded to consider the charge-conjugate state $ D_{s} $ and $ D $ mesons as well. In such case, it is necessary to consider additional new channels such as %$ \Lambda_{c}K^{+}-pD_{s} $  
$ D_{s}N\rightarrow K\Lambda_{c} $
~\cite{YasuiPRC2017}, which is tackled here.

Therefore, a determination of the scattering parameters of systems
involving charm hadron %like $\textrm{D}$ and/or $\textrm{D}^{*}$ mesons
 are essential. % to describe numerous observed exotic states in QCD. 
Since the nature of these interactions is  due to non-perturbative QCD, the best tool to study them is
  lattice QCD simulations that is based on the first-principles calculations of QCD.   
Today this is possible by the modern high performance computing facilities together
 progress in theoretical method like HAL QCD~\cite{AokiFP2020} and non-relativistic effective field theories~\cite{Epelbaum2009,BRAMBILLA20201}. They  have equipped us to explore new forms of matter.

Motivating by the above discussions, it is desirable to determine the coupled channel $ \Lambda_{c}K^{+}-pD_{s} $ interactions in $I\left(J^{p}\right)=1/2^{-}$ state ($\Lambda_{c}$ : charmed baryon$\left(udc\right)$,
$K^{+}$ : Kaon$\left(u\bar{s}\right)$, $D_{s}$: Strange $\:D\:$ meson
: $\left(c\bar{s}\right)$ and $p$ is proton $\left(uud\right)$), by lattice QCD simulations.
In the recent years, an approach to investigate hadron interactions
in lattice QCD has been proposed by the HAL QCD Collaboration~\cite{IshiiPRL2007,AokiPTEP2012}.
One of the advantages of the HAL QCD method is that it can be extended
straightforwardly to the case of inelastic scatterings. The extended
method, namely coupled-channel HAL QCD method~\cite{AokiPrd2013},
has been applied to the hyperon-baryon(hyperon) systems~\cite{SasakiPTEP2015,sasaki2020},
 charmed baryon-(charmed) baryon systems~\cite{MIYAMOTO2018113,YanPRL2021},  charmed tetraquarks states ~\cite{IkedaPLB2014,IkedaPRL2016,YanPRL2023}, resonance states~\cite{AkahoshiPRD2021} and meson-baryon bound states~\cite{ikeda2011structure,MurakamiPTEP2023}. The calculated
scattering amplitude  from obtained potentials can be used
to compare the scattering observables with experimental data~\cite{IkedaPRL2016}.

Here, we consider the inelastic regions for the $ pD_{s} $ scattering
by using coupled-channel HAL QCD method. In particular, we focus on
the S-wave $ \Lambda_{c}K^{+}-pD_{s} $ and   
%($\Lambda_{c}$ : charmed baryon$\left(udc\right)$, $K^{+}$ : kaon$\left(u\bar{s}\right)$, $D_{s}$: Strange $\:D\:$ meson : $\left(c\bar{s}\right)$ and $p$ is proton $\left(uud\right)$) system in $I\left(J^{p}\right)=1/2^{-}$ state. 
%For better understanding properties of the S-wave $\Lambda_{c}K^{+}-pD_{s}$ interaction, we
calculate the scattering observables from obtained potentials in the
infinite volume, such as phase shifts for $\Lambda_{c}K^{+}$ and
$ pD_{s} $ systems and the inelasticity of the scatterings.

This paper is organized as follows. In section~\ref{sec:HAL-method-}, we
review the coupled-channel approach to the baryon-meson interactions
by the HAL QCD method in lattice QCD. In section~\ref{sec:Numerical-setup},
the numerical setup on the lattice and definitions of baryon and meson
operators are summarized. We present our results on the coupled-channel
potential for the $\Lambda_{c}K^{+}-pD_{s}$ system in the S-wave
$J^{p}=1/2^{-}$ state, the phase shift and scattering length by solving Schr{\"{o}}dinger equation with the extracted potential in section~\ref{sec:Numerical-results}. And
finally, section~\ref{sec:Summary-and-conclusion} is devoted
to summary and conclusion.

\section{HAL method for coupled channel\label{sec:HAL-method-}}
Here,  we describe briefly the coupled-channel HAL QCD method~\cite{AokiPrd2013},
which will be applied to the $\Lambda_{c}K^{+}-pD_{s}$ system. A
key quantity in the HAL QCD method is the equal-time Nambu-Bethe-Salpeter
(NBS) wave function which encodes information of scattering amplitude
in its asymptotic behaviour. In the center-of-mass frame, the NBS
wave function of baryon-meson at Euclidean time $t$ with the total
energy $ W $ is defined by

\begin{equation}
\Psi_{C}^{\left(W\right)}\left(\vec{r}\right)e^{-Wt}=\frac{1}{\sqrt{Z_{C_{1}}}\sqrt{Z_{C_{2}}}}\sum_{\vec{x}}\left\langle 0\left|B_{C_{1}}\left(\vec{r}+\vec{x},t\right)\phi_{C_{2}}\left(\vec{x},t\right)\right|W\right\rangle ,\label{eq:nbs}
\end{equation}
where the index $C$ denote the flavor channel $\left(C=\Lambda_{c}K^{+},pD_{s}\right)$,
and $B_{C}\left(\phi_{C}\right)$ is the local interpolating operator
for the baryon(meson) $C_{i}$ with its renormalization factor $Z_{C_{i}}$. In the case of $C=\Lambda_{c}K^{+}$, for instance, $C_{1}=\Lambda_{c}$
and $C_{2}=K^{+}$. The $\left|W\right\rangle $ stands for a QCD
asymptotic in-state at the total energy of $W$. From the NBS wave
functions, we define the energy independent non-local potentials through
the following coupled-channel Schr{\"{o}}dinger equation,

\begin{equation}
\left[E_{C}-\left(H_{0}\right)_{C}\right]\Psi_{C}^{\left(W\right)}\left(\vec{r}\right)=\sum_{C^{\prime}}\int d^{3}\vec{r}^{\prime}U_{C}^{C^{\prime}}\left(\vec{r},\vec{r}^{\prime}\right)\Psi_{C^{\prime}}^{\left(W\right)}\left(\vec{r}^{\prime}\right),\label{eq:ccshe}
\end{equation}
where $\left(H_{0}\right)_{C}=-{\nabla^{2}}/{2\mu_{C}}$ with
the reduced mass $\mu_{C}={m_{C_{1}}m_{C_{2}}}/{\left(m_{C_{1}}+m_{C_{2}}\right)}$
and $E_{C}={k_{C}^{2}}/{2\mu_{C}}$. The relative momentum
$k_{C}$ is determined from the total energy $W=\sqrt{k_{C}^{2}+m_{C_{1}}}+\sqrt{k_{C}^{2}+m_{C_{2}}}$.
By definition, the non-local potential $U\left(\vec{r},\vec{r}^{\prime}\right)$
is faithful to the QCD S-matrix unless new channel opens. In order
to handle the non-locality of the potentials, we introduce the derivative
expansion $U\left(\vec{r},\vec{r}^{\prime}\right)=\left(V_{LO}\left(\vec{r}\right)+V_{NLO}\left(\vec{r}\right)+...\right)\delta\left(\vec{r},\vec{r}^{\prime}\right)$
, where the $N^{n}LO$ term is of $\mathcal{O}\left(\vec{\nabla}^{n}\right)$.
The leading- order potential matrix is extracted by using the NBS
wave functions as 
\begin{equation}
\left(\begin{array}{cc}
V_{\Lambda_{c}K^{+}}^{\Lambda_{c}K^{+}} & V_{\Lambda_{c}K^{+}}^{pD_{s}}\\
V_{pD_{s}}^{\Lambda_{c}K^{+}} & V_{pD_{s}}^{pD_{s}}
\end{array}\right)=\left(\begin{array}{cc}
K_{\Lambda_{c}K^{+}}^{W_{1}} & K_{\Lambda_{c}K^{+}}^{W_{2}}\\
K_{pD_{s}}^{W_{1}} & K_{pD_{s}}^{W_{2}}
\end{array}\right)\left(\begin{array}{cc}
\Psi_{\Lambda_{c}K^{+}}^{W_{1}} & \Psi_{\Lambda_{c}K^{+}}^{W_{2}}\\
\Psi_{pD_{s}}^{W_{1}} & \Psi_{pD_{s}}^{W_{2}}
\end{array}\right)^{-1},\label{eq:lopm}
\end{equation}
where, $K_{C}^{W}\left(\vec{r}\right)=\left[E_{C}-\left(H_{0}\right)_{C}\right]\Psi_{C}^{\left(W\right)}\left(\vec{r}\right)$. 

In lattice QCD, the NBS wave functions can be extracted from the baryon-meson
four-point correlation function given by
\begin{eqnarray}
G_{C}^{C^{\prime}}\left(\vec{r},t-t_{0}\right) & = & \sum_{\vec{x}}\left\langle 0\left|B_{C_{1}}\left(\vec{r}+\vec{x},t\right)\phi_{C_{2}}\left(\vec{x},t\right)
\overline{\mathfrak{J}^{C^{\prime}}}\left(t_{0}\right)\right|0\right\rangle \nonumber \\
& = & \sum_{n}\sqrt{Z_{C_{1}}}\sqrt{Z_{C_{2}}}\Psi_{C}^{\left(W_{n}\right)}\left(\vec{r}\right)e^{-W_{n}\left(t-t_{0}\right)}A_{n}^{C^{\prime}}+\ldots,\label{eq:4pt}
\end{eqnarray}
with constant $A_{n}^{C^{\prime}}=\left\langle W_{n}\left|\overline{\mathfrak{J}^{C^{\prime}}}\left(t_{0}\right)\right|0\right\rangle $,
where $\overline{\mathfrak{J}^{C^{\prime}}}\left(t_{0}\right)$ stands
for the source operator for $C^{\prime}$ which creates baryon-meson
states. The ellipses denote inelastic contributions coming from channels
above $C$ and $C^{\prime}$. 
In fact, for large time the  four-point correlation function in Eq~\eqref{eq:4pt} is dominated by the ground state NBS
wave function.
Practically, however, the accurate determination
of potentials have some difficulties, first, it is to figure out the ground
state domination, since $t-t_{0}$ can not be taken large enough due
to statistical noises of the baryon-meson four-point correlation function~\cite{IritaniJHEP2016,IritaniPRD2017},
and second, in order to solve Eq.~\eqref{eq:lopm}, we need not only
the ground state of NBS wave functions but also the first excited
state of NBS wave functions, which are difficult to isolate each other
due to the same reasons as former. The improved method to overcome
these practical difficulties, has been proposed in Ref.~\cite{IshiiPLB2012}
in the case of the single channel and extended to the coupled-channel
case in Refs.~\cite{AokiPTEP2012,AokiPrd2013}. Let us consider the
normalized baryon-meson four-point correlation function

\begin{equation}
R_{C}^{C^{\prime}}\left(\vec{r},t-t_{0}\right)\equiv G_{C}^{C^{\prime}}\left(\vec{r},t-t_{0}\right)/\exp\left[-\left(m_{C_{1}}+m_{C_{2}}\right)\left(t-t_{0}\right)\right],\label{eq:n4pt}
\end{equation}
this satisfies
\begin{equation}
\left[\left(\frac{1+3\delta_{C}^{2}}{8\mu_{C}}\right)\frac{\partial^{2}}{\partial t^{2}}-\frac{\partial}{\partial t}-\left(H_{0}\right)_{C}\right]R_{C}^{C^{\prime}}\left(\vec{r},t-t_{0}\right) \label{eq:cise}
 \end{equation}
$$   =\sum_{C^{\prime\prime}}\int\mathop{d^{3}\vec{r}^{\prime}\Delta_{C}^{C^{\prime\prime}}U_{C}^{C^{\prime\prime}}\left(\vec{r},\vec{r}^{\prime}\right)R_{C^{\prime\prime}}^{C^{\prime}}\left(\vec{r}^{\prime},t-t_{0}\right),}
$$
for a moderately large $t-t_{0}$ where inelastic contributions from
channels other than $\Lambda_{c}K^{+}$ and $pD_{s}$ can be neglected (Of course Eq.~\eqref{eq:cise} is not exact). 
$\delta_{C}$ and $\Delta_{C}^{C^{\prime}}$ in Eq.~\eqref{eq:cise}
are defined by
\begin{equation}
\delta_{C}=\left(m_{C_{1}}-m_{C_{1}}\right)/\left(m_{C_{1}}+m_{C_{1}}\right),
\end{equation}

\begin{equation}
\Delta_{C}^{C^{\prime\prime}}=\sqrt{\left(Z_{C_{1}}Z_{C_{2}}\right) / \left(Z_{C_{1}^{\prime\prime}}Z_{C_{2}^{\prime\prime}}\right)}\exp\left[-\left(m_{C_{1}^{\prime\prime}}+m_{C_{2}^{\prime\prime}}-m_{C_{1}}-m_{C_{2}}\right)\left(t-t_{0}\right)\right].
\end{equation}

We can then extract the potential matrix at the leading-order of the
derivative expansion as

\begin{equation}
\left(\begin{array}{cc}
\tilde{V}_{\Lambda_{c}K^{+}}^{\Lambda_{c}K^{+}}\left(\vec{r}\right) & \tilde{V}_{\Lambda_{c}K^{+}}^{pD_{s}}\left(\vec{r}\right)\\
\tilde{V}_{pD_{s}}^{\Lambda_{c}K^{+}}\left(\vec{r}\right) & \tilde{V}_{pD_{s}}^{pD_{s}}\left(\vec{r}\right)
\end{array}\right)=
\end{equation}
$$
\left(\begin{array}{cc}
\mathcal{K}_{\Lambda_{c}K^{+}}^{\Lambda_{c}K^{+}}\left(\vec{r},t-t_{0}\right) & \mathcal{K}_{\Lambda_{c}K^{+}}^{pD_{s}}\left(\vec{r},t-t_{0}\right)\\
\mathcal{K}_{pD_{s}}^{\Lambda_{c}K^{+}}\left(\vec{r},t-t_{0}\right) & \mathcal{K}_{pD_{s}}^{pD_{s}}\left(\vec{r},t-t_{0}\right)
\end{array}\right)\left(\begin{array}{cc}
R_{\Lambda_{c}K^{+}}^{\Lambda_{c}K^{+}}\left(\vec{r},t-t_{0}\right) & R_{\Lambda_{c}K^{+}}^{pD_{s}}\left(\vec{r},t-t_{0}\right)\\
R_{pD_{s}}^{\Lambda_{c}K^{+}}\left(\vec{r},t-t_{0}\right) & R_{pD_{s}}^{pD_{s}}\left(\vec{r},t-t_{0}\right)
\end{array}\right)^{-1},
$$
where $\tilde{V}_{C}^{C^{\prime}}\left(\vec{r}\right)=\Delta_{C}^{C^{\prime}}V_{C}^{C^{\prime}}\left(\vec{r}\right)$
and $\mathcal{K}_{C}^{C^{\prime}}\left(\vec{r},t-t_{0}\right)\equiv$
(l.h.s of Eq.~\eqref{eq:cise}).

\section{Numerical Lattice setup\label{sec:Numerical-setup}}
%by the CP-PACS and JLQCD collaborations[31]  … . These configurations are available through Japan Lattice Data Grid (JLDG)[32].
  We applied the 3-flavor full QCD gauge configurations that are generated by the CP-PACS and JLQCD 
Collaborations~\cite{CP-PACS/JLQCD}
with the renormalization-group-improved gauge action and the non-perturbatively
$\mathcal{O}\left(a\right)$-improved Wilson quark action $ \left(c_{SW}=1.7610\right) $ at $\beta=6/g^{2}=1.83$
(corresponding lattice spacing in the physical unit, $a=0.1209$ fm
~\cite{IshikawaPRD2008}) on an $L^{3}\times T=16^{3}\times32$ lattice, that corresponds to  $\left(1.93\right)^{3}\times3.87$ fm in the physical unit. These configurations are available through Japan Lattice Data Grid (JLDG)~\cite{JLDG}. 
The hopping parameters of the configuration
set correspond to the flavor $ \textrm{SU}\left(3\right) $ symmetric point
are $\kappa_{u,d}=\kappa_{s}=0.13710$.
%\textcolor{red}{ In the case of the charm quark, the relativistic heavy quark (RHQ) action~\cite{AokiPTP2003} is employed to avoid the leading $\mathcal{O}\left(\left(m_{Q}a\right)^{n}\right)$ and the next-to-leading $\mathcal{O}\left(\left(m_{Q}a\right)^{n}\left(a\Lambda_{QCD}\right)\right)$ discretization errors due to the charm quark mass $m_{Q}$. We applied the RHQ parameters defined in Ref.~\cite{NamekawaPRD} in such a way to reproduce the physical value of the mass and the relativistic dispersion relation for the charmonium in the spin-averaged $ 1S $ state.}
%\textcolor{blue}{
	In the case of the charm quark, also the non-perturbatively
	$\mathcal{O}\left(a\right)$-improved Wilson quark action $ \left(c_{SW}=1.7610\right) $ is used. The charm
	quark propagators are calculated at $ \kappa_{c} = 0.12240 $ in (partial) quenched QCD~\cite{CAN2013703}.
%}
%The number of gauge configurations and various hadron masses calculated
%in this work are summarized in Table \ref{tab:Hadron-masses-in}.

We use $700$ gauge configurations, and the quark propagators are calculated for the wall source at $t_{0}$. 
The periodic boundary conditions are imposed in
the three spacial directions, while Dirichlet boundary conditions
are taken for the  temporal direction at $t=16+t_{0}$.
In order to increase the statistics, beside averaging over forward/backward propagations, the wall source is placed at $32$ different
place of $t_{0}$ on each configuration. 
In this work, the jackknife method is used to estimate statistical error with a bin
size of $20$ configurations. 

For the local interpolating operators in Eq.~\eqref{eq:4pt}, we use
following form for proton and $\Lambda_{c}$ as
\begin{equation}
B_{\alpha}\left(x\right)=\varepsilon_{ijk}\left[q_{i}^{T}\left(x\right)C\gamma_{5}q_{j}\left(x\right)\right]q_{k,\alpha}\left(x\right),
\end{equation}
And for mesons,
\begin{equation}
\phi\left(x\right)=\bar{q}\left(x\right)\gamma_{5}q\left(x\right),
\end{equation}
where $x=\left(\vec{x},t\right)$, and $i,j,k$ are color indices.
$C$ is the charge conjugation matrix defined by $C=\gamma_{2}\gamma_{4}$,
and $q=u,d,s,c$ stands for quark operators for up-, down-, strange- and charm-
quarks, respectively. Flavor structures of a p, $\Lambda_{c},K^{+}$ and $D_{s}$
are given by
\begin{equation}
\begin{array}{cc}
p=\left[ud\right]u, & \Lambda_{c}^{+}=\frac{1}{\sqrt{6}}\left(\left[cd\right]u+\left[uc\right]d-2\left[du\right]c\right)\\
K^{+}=u\bar{s}, & D_{S}^{+}=c\bar{s}.
\end{array}
\end{equation}
%\begin{table}
%	\begin{center}
%			\caption{Hadron masses in unit of {[}MeV{]}.\label{tab:Hadron-masses-in}}
%		\begin{tabular}{|c|c|c|c|c|}
%			\hline 
%			confs. & $\Lambda_{c}$ & $D_{s}$ & $p$ & $K^{+}$\tabularnewline
%			\hline 
%			\hline 
%			\textcolor{red}{\#\#\#} & 2583$\pm$2.4 & 1825$\pm$0.9 & 1810$\pm$3.2 & 910$\pm$1\tabularnewline
%			\hline 
%		\end{tabular}	
%	\end{center}
%\end{table}

\section{Numerical results\label{sec:Numerical-results}}

\subsection{Effective mass and renormalization factor}
\begin{figure} [htp]
	\begin{center}
		\includegraphics[scale=0.7]{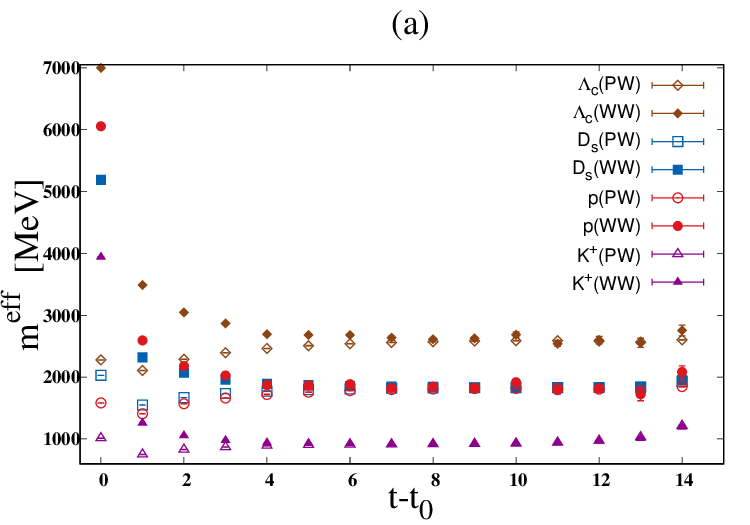}\includegraphics[scale=0.7]{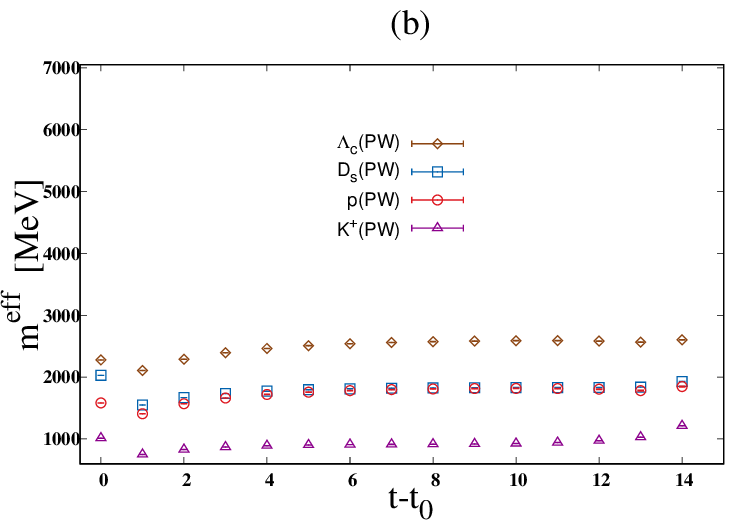}
		\caption{Effective mass plot of $\Lambda_{c}, D_{s}, p$ and $K^{+}$ for (a)  wall-wall (WW) and  point-wall (PW),  in addition, to have better comparison in the plot (b) only point-wall correlators are shown. \label{fig:ems}}
	\end{center}
\end{figure}

We show the effective mass plots of the temporal two-point correlators
of $\Lambda_{c},D_{s},p$ and $K^{+}$ in Fig.~\ref{fig:ems} both for the point-sink
and wall-source (point-wall) and for the wall-sink and wall-source
(wall-wall). To obtain the mass of hadron, $ m_{H} $ and the overlap parameters $ a_{PW} $
and $ a_{WW} $ for the point-wall and wall-wall correlators, we perform
single exponential fit analysis of the point-wall and the wall-wall
temporal two-point correlators by employing the functional form
\begin{eqnarray}
\begin{array}{cc}
C_{PW}\left(t\right)\simeq a_{PW}\exp\left(-m_{H}t\right), & C_{WW}\left(t\right)\simeq a_{WW}\exp\left(-m_{H}t\right), \label{eq:cpw_cww} \end{array}
\end{eqnarray}
 the $ Z_{H} $ factor for a hadron is defined by
\begin{equation}
\sqrt{Z_{H}}=\sqrt{2 m_{H}}\frac{a_{PW}}{\sqrt{a_{WW}}}. \label{eq:z-fac}
\end{equation}
 The $ Z_{H} $  can be calculated numerically through fitting the hadron correlators with exponential function.
%\textcolor{red}{a relation between ($ a_{PW}, a_{WW} $) and (Z’s should be given here.}

The overlap parameters are used to obtain the $\Lambda_{c}K^{+}-pD_{s}$
coupled channel potentials. They contribute to the $\Lambda_{c}K^{+}-pD_{s}$
coupled channel potential in the combination $\sqrt{\left(Z_{\Lambda_{c}}Z_{K^{+}}\right)/\left(Z_{p}Z_{D_{s}}\right)}$.
Here, $Z_{\Lambda_{c}}, Z_{K^{+}}, Z_{p}$ and $Z_{D_{s}}$ denote the
$Z$ factors of local composite hadron operators for $\Lambda_{c}, K^{+}, p$
and $ D_{s} $ hadrons, respectively. 
%, which appear in the limit $\psi\left(x\right)\rightarrow Z^{1/2}\psi_{out}$ as $x_{0}\rightarrow+\infty$ where $\psi\left(x\right)$ and $\psi_{out}$ denote local composite hadron operators and the corresponding asymptotic fields.
 The identification plateau regions and the results for the hadron masses are
given in Table~\ref{tab:Z-fac}, whereas these values leads to $\sqrt{\left(Z_{\Lambda_{c}}Z_{K^{+}}\right)/\left(Z_{p}Z_{D_{s}}\right)} =1.22 $.

\begin{table}
	\begin{center}
		\caption{Hadron masses in unit of  MeV. Statistical errors are shown in the parentheses. \label{tab:Z-fac}}%\label{tab:Hadron-masses-in}}
		%\rowcolors{1}{green!80!yellow!50}{green!70!yellow!40}
		%\textcolor{red}{definitions of $ Z $ and $ Z_P $ should be given.}
		The $ Z_{H} $ is defined in Eq.~\eqref{eq:z-fac} and the $ Z_{p} $ is the Z-factor of proton.
		\begin{tabular}{c c c c c c}
			\hline 
			Hadron &  $\sqrt{Z_{H}}$ &  $\sqrt{\frac{Z_{H}}{Z_{p}}}$ &  $m_{H}\left[\textrm{MeV}\right]$ & Fit Range & $m_{exp}[\textrm{MeV}]$\tabularnewline
			\hline 
			\hline 
			$\pi$   &             &           &  $ 872(3) $ & $ 6-10 $ & $ 135 $\tabularnewline
			$K$   &  $ 1.478(7) $ & $ 6.234 $ &  $ 910(1) $ & $ 5-10 $ & $ 493.67 $\tabularnewline
			
			$p$   & $ 0.2371(2) $ & $ 1.000 $ &  $ 1810(3) $ &$ 7-12 $ &$ 938.27 $\tabularnewline
			
			$D_{s}$& $ 1.81(1)  $ & $ 7.662 $ &  $ 1825(1) $ &$ 7-11 $ &$ 1968.30 $\tabularnewline
			
			$\Lambda_{c}$ &  $ 0.3569(9) $ & $ 1.505 $ &  $ 2583(3) $ &$ 8-12 $ &$ 2286.46 $\tabularnewline
			\hline 
		\end{tabular}
	\end{center}
\end{table}

\subsection{Time dependence}
\begin{figure}
	\begin{center}
		\includegraphics[scale=0.7]{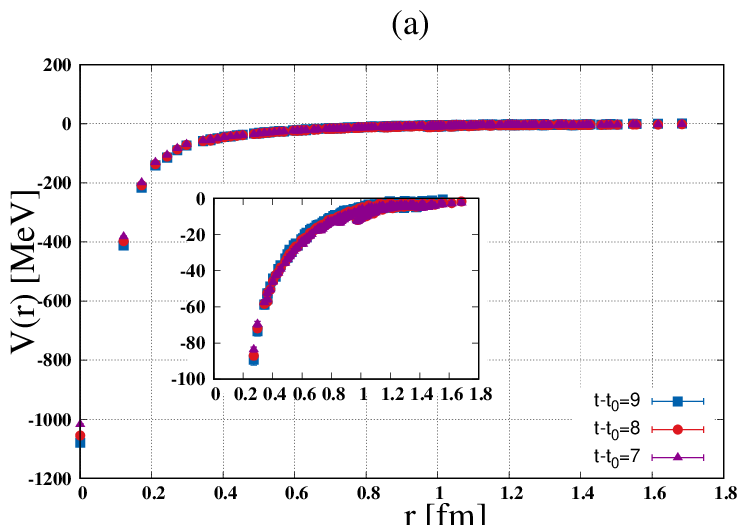}\includegraphics[scale=0.7]{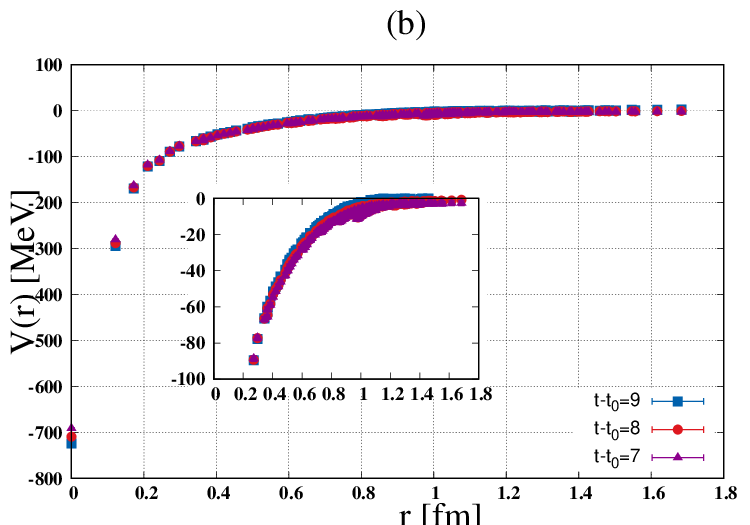}
		\caption{Time dependences of the diagonal potentials matrix, panel (a) for
			$V_{\Lambda_{c}K^{+}}^{\Lambda_{c}K^{+}}$ and panel (b) for $V_{pD_{s}}^{pD_{s}}$
			%in the\textcolor{red}{{} ${\color{black}I\left(J^{p}\right)=1/2\left(1/2^{-}\right)}$} stateas
			 as function of $r$ at $t-t_{0}=7$ (darkmagenta), $8$ (red)
			and $9$ (blue). \label{fig:t_dep}}
	\end{center}
\end{figure}

First, we investigate the time dependence of the diagonal potentials.
The time interval $ t-t_{0}=8 \pm 1 $ is selected to suppress contribution from
higher excited states at smaller $ t $ and
simultaneously to avoid large statistical errors at larger $ t $~\cite{SasakiPTEP2015}.
Fig.~\ref{fig:t_dep} shows $V_{\Lambda_{c}K^{+}}^{\Lambda_{c}K^{+}}$
and $V_{pD_{s}}^{pD_{s}}$ at three values of $ t-t_{0}=7,8,9$. 
Based on the time-dependent HAL QCD method~\cite{IshiiPLB2012},
within statistical errors, no significant $ t-t_{0} $ dependence is observed
for these potentials showing that $ t-t_{0}=8 $ is large enough to
suppress inelastic contributions and that higher-order contributions
in the derivative expansion are negligible.

\subsection{Potentials\label{subsec:Potentials}}

\begin{figure}
	\begin{center}
		
		\includegraphics[scale=0.7]{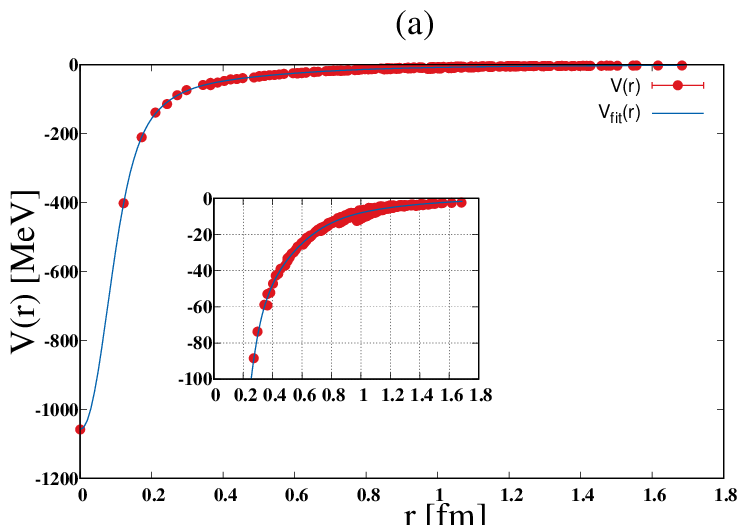}\includegraphics[scale=0.7]{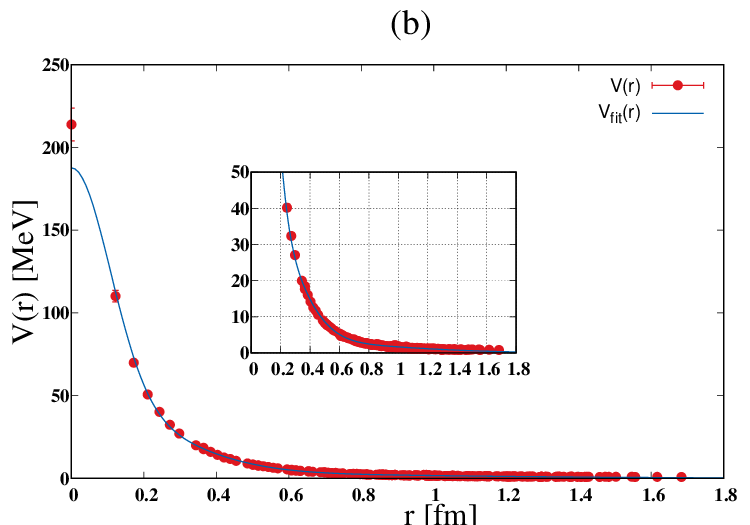} 
		
		\includegraphics[scale=0.7]{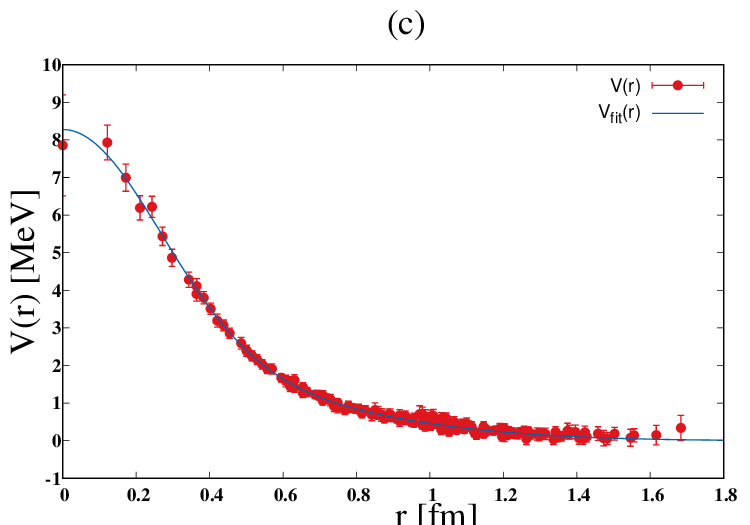}\includegraphics[scale=0.7]{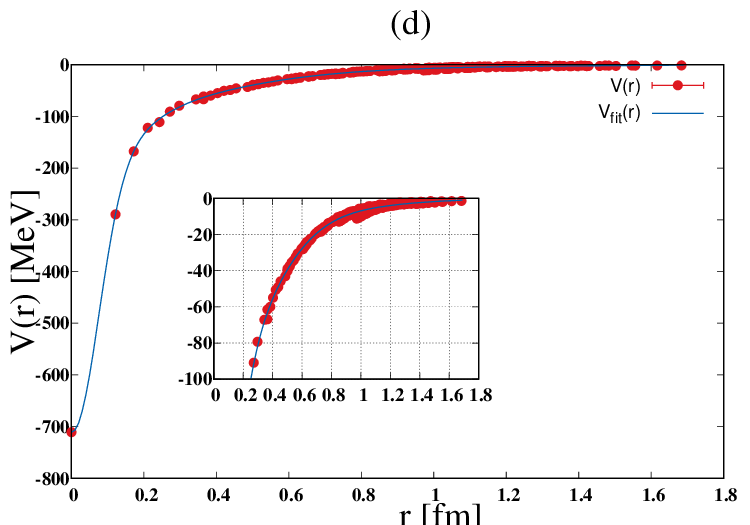}
		\caption{The S-wave $\Lambda_{c}K^{+}-pD_{s}$ coupled-channel potential matrix (red circle). The solid blue line is the corresponding fitted function.
			%in the\textcolor{red}{{} ${\color{black}J^{p}=1/2^{-}}$} state.
			 The
			time separation between sink and source is $t-t_{0}=8$. Panels show
			elements of the potential matrix, (a) $ V_{\Lambda_{c}K^{+}}^{\Lambda_{c}K^{+}}$, (b)  $V_{\Lambda_{c}K^{+}}^{pD_{s}}$, 
			(c) $ V_{pD_{s}}^{\Lambda_{c}K^{+}} $ and (d) $V_{pD_{s}}^{pD_{s}}$, respectively.\label{fig:pots}}
	\end{center}
	
\end{figure}

The numerical results of the S-wave $\Lambda_{c}K^{+}-pD_{s}$ coupled-channel
potential matrix % in the \textcolor{red}{${\color{black}I\left(J^{p}\right)=1/2\left(1/2^{-}\right)}$} state,
 are presented in Fig~\ref{fig:pots}. This figure  shows attractions at
all distances without repulsive core for diagonal elements of the
potential matrix $ V_{\Lambda_{c}K^{+}}^{\Lambda_{c}K^{+}} $ and $ V_{pD_{s}}^{pD_{s}} $,
and the attraction at short distance $\left(r<0.5\right)$ fm in the
$\Lambda_{c}K^{+}$ channel is stronger than in the $pD_{s}$ channel.
These (diagonal and off-diagonal) potentials, almost vanish at $ r>1 $ fm.

\subsection{Hermiticity}\label{subsec:hermicity}
Hermiticity of the potential matrix is a sufficient condition for
the probability conservation, though it is not a necessary condition.
Furthermore, the Hermiticity of the potential matrix is not automatically guaranteed in the definition of the
coupled-channel potential matrix in Eq.~\eqref{eq:ccshe}.
%We investigate the Hermiticity of the potential matrix, $V_{b}^{a}=V_{a}^{b}$ , since it is not automatically guaranteed in the definition of the coupled-channel potential matrix in Eq.~\eqref{eq:ccshe}. As in the case of the diagonal parts, we confirm that the off-diagonal parts of the potential matrix show no significant $t-t_{0}$ dependence, so we take the results at $t-t_{0}=8$ in our analysis. We introduce a Hermiticity measure 
%\begin{equation}
%\delta V_{a-b}\equiv2\frac{\left(V_{b}^{a}-V_{a}^{b}\right)}{\left(V_{b}^{a}+V_{a}^{b}\right)}, \label{eq:hermicity}
%\end{equation}
%to see the relative magnitude of the Hermiticiy violation of the potential matrix. Figure~\ref{fig:Violation-of-Hermiticity} presents the values of criterion  $\delta V_{\Lambda_{c}K^{+}-pD_{s}}$ in Eq.~\eqref{eq:hermicity}.
 %in the\textcolor{red}{{} ${\color{black}I\left(J^{p}\right)=1/2\left(1/2^{-}\right)}$} channel.
  From Fig.~\ref{fig:pots} it is apparent that the Hermiticity is largely broken, i.e. the off-diagonal parts
  of the potential (panel (b) and (c) in Fig.~\ref{fig:pots}) matrix are different. 
 This may indicate that it could be better  to include the $ pD_{s}^{\ast} $ channel into our calculation.
To treat this broken Hermiticity, we have done further analysis. %subsection~\ref{subsec:more_analysis}. 
%\begin{figure}
%	\begin{center}
%		\includegraphics[scale=0.7]{deltaV_LamCKP_ProtDs_T8}\caption{Violation of Hermiticity, i.e.,
%		% in the \textcolor{red}{{} ${\color{black}I\left(J^{p}\right)=1/2\left(1/2^{-}\right)}$} 	channel 
%		$\delta V_{\Lambda_{c}K^{+}-pD_{s}}$ in Eq.~\eqref{eq:hermicity}
%		at $ t-t_{0}=8 $.\label{fig:Violation-of-Hermiticity}}
%	\end{center}
%\end{figure}
 Solving the coupled-channel Schr{\"{o}}dinger equation with non-Hermitian potential is problematic because the unitarity of the S-matrix is not guaranteed. Therefore, 
 it is common to apply Hermitian potential in phenomenological studies in nuclear physics. Since this broken Hermiticity is happened at short distances, the low energy scattering observables does not suffer from difference between $ V_{\Lambda_{c}K^{+}}^{pD_{s}} $ and $ V_{pD_{s}}^{\Lambda_{c}K^{+} } $.
 In order to check this the coupled channel potential is made  Hermitian by taking one of them for the off-diagonal part or their average,
 
 \begin{equation}
 \bar{V}_{\Lambda_{c}K^{+}}^{pD_{s}}=\left(V_{\Lambda_{c}K^{+}}^{pD_{s}}+V_{pD_{s}}^{\Lambda_{c}K^{+}}\right)/2. \label{eq:vbar}
 \end{equation}
 In Appendix~\ref{subsec:more_analysis}, it is shown clearly that
 the scattering phase shifts of $ {\Lambda_{c}K} $ and $ {pD_{s}} $ for the above $ 3 $ Hermitian potential cases are almost same within the statistical errors. Thus hereafter  we consider the $ \bar{V}_{\Lambda_{c}K^{+}}^{pD_{s}} $ in our calculations. 

\section{Analytic forms of $ {\Lambda_{c}K^{+}} $ and $ {pD_{s}} $ potentials}
In order to use the LQCD potential in phenomenological investigations, it is useful to fit them with some functions. 
Accordingly for the diagonal (D) $ {\Lambda_{c}K^{+}}-{\Lambda_{c}K^{+}} $
 $ \left(V_{\Lambda_{c}K^{+}}^{\Lambda_{c}K^{+}}\right) $, 
 $ {pD_{s}}-{pD_{s}} $ $ \left(V_{pD_{s}}^{pD_{s}}\right) $ potentials as presented in Fig.~\ref{fig:pots}, the following fit function $ V_{D}\left(r\right) $ is considered~\cite{etminan2014,Iritani2019prb}, 
\begin{equation}
V_{\textrm{D}}\left(r\right)={\displaystyle \sum_{i=1}^{3}\alpha_{i}e^{-\beta_{i}r^{2}}}+\lambda\left[\mathcal{Y}\left(\rho,m_{\pi},r\right)\right]^{2}, \label{eq:V_D}
\end{equation}
%The form of analytic fit function for $\bar{V}_{\Lambda_{c}K^{+}}^{pD_{s}} $ in Eq.~\eqref{eq:vbar} is same as for Eq.~\eqref{eq:V_OD}. The results of the fit and the corresponding parameters are summarized in Fig.~\ref{fig:vbar} and Table~\ref{tab:vbar}, respectively. 

\begin{figure}
	\begin{center}
		\includegraphics[scale=0.7]{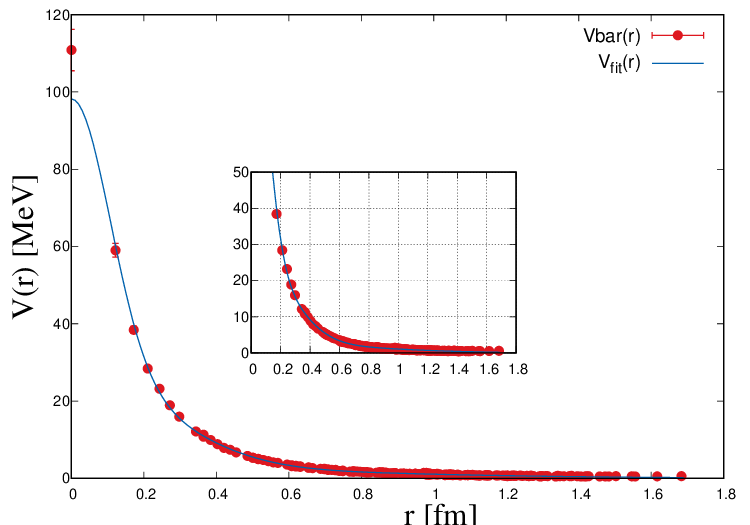}\caption{ Hermitian (average) potential, $\bar{V}_{\Lambda_{c}K^{+}}^{pD_{s}} $ in Eq.~\eqref{eq:vbar} and its fit function at $t-t_{0}=8$.\label{fig:vbar}}
	\end{center}
\end{figure}

And similarly for the off-diagonal(OD) 
%$ {\Lambda_{c}K^{+}}-{pD_{s}} $ 
%$ \left(V_{\Lambda_{c}K^{+}}^{pD_{s}}\right) $, 
%$  {pD_{s}}-{\Lambda_{c}K^{+}}$ 
%$ \left(V_{pD_{s}}^{\Lambda_{c}K^{+}}\right) $ 
 $\bar{V}_{\Lambda_{c}K^{+}}^{pD_{s}} $ as presented in Fig.~\ref{fig:vbar}, the following analytic form
 $ V_{\textrm{OD}}\left(r\right) $ is selected~\cite{Gongyo2018},
%\begin{equation}
%V_{pD_{s}}^{pD_{s}}\left(r\right)={\displaystyle \alpha e^{-\beta r^{2}}}+\lambda %\left[\mathcal{Y}\left(\rho_{1},\rho_{2},r\right)\right]^{2},\label{eq:V_pD-pD}
%\end{equation}
\begin{equation}
V_{\textrm{OD}}\left(r\right)={\displaystyle \sum_{i=1}^{3}\alpha_{i}e^{-\beta_{i}r^{2}}},\label{eq:V_OD}
\end{equation}
where $ \mathcal{Y} $ is the Yukawa function with a form factor,
\begin{equation}
\mathcal{Y}\left(\rho,m_{\pi},r\right)\equiv\left(1-e^{-\rho r^{2}}\right)\frac{e^{-m_{\pi}r}}{r}.
\end{equation}
In the Eq.~\eqref{eq:V_D},  the $ \mathcal{Y} $ form, at medium and long range distances is motivated by the picture of two-pion exchange. The Gauss form factors in the Eq.~\eqref{eq:V_D} and~\eqref{eq:V_OD} describe the short range part of the potentials. Note that the pion mass $ m_{\pi} $ are fixed to be the measured values on the lattice about $ 872 $ MeV.

The results of fit and
corresponding parameters are given in Table~\ref{tab:V_LK-LK} for 
$ V_{\Lambda_{c}K^{+}}^{\Lambda_{c}K^{+}} $, Table~\ref{tab:V_pDs-pDs} for $ V_{pD_{s}}^{pD_{s}} $
%, Table~\ref{tab:V_LK-pD} for $ V_{\Lambda_{c}K^{+}}^{pD_{s}} $  and Table~\ref{tab:V_pD-LK} for $ V_{pD_{s}}^{\Lambda_{c}K^{+}} $
 and Table~\ref{tab:vbar} for $\bar{V}_{\Lambda_{c}K^{+}}^{pD_{s}} $ at three different values $ t-t_{0} = 7, 8, 9 $.

%%%%%%%%%%%%%%%%%%%%%%%%
\begin{table}
	\caption{
		Fitted parameters in Eq.~\eqref{eq:V_D} for $ V_{\Lambda_{c}K^{+}}^{\Lambda_{c}K^{+}} $ with statistical errors using the data at $ r<1.7 $ fm.
		$ \alpha_{i},\beta_{i} $ and $ \rho $ are given in units of 
		$ \left[\textrm{MeV}\right],\left[\textrm{fm}^{-2}\right] $ and
		$ \left[\textrm{fm}^{-2}\right] $, respectively. 
		%\textcolor{red}{The values of 	$ \chi^{2}/{d.o.f} $ with $ d.o.f=214 $ are $ 1.30(40),0.76(18) $ and $ 0.74(30) $ for $ t-t_{0}=7,8 $ and $ 9 $, respectively.}
	}
	\centering
	%	\begin{ruledtabular}
	\begin{tabular}{ccccccccccccc}
		\hline
		&	  \multicolumn{2}{c}{Gauss-1}    && \multicolumn{2}{c}{Gauss-2}    &&  \multicolumn{2}{c}{Gauss-3} &&  \multicolumn{2}{c}{$ {\textrm{Yukawa}}^{2} $}\\ \cline{2-3} \cline{5-6} \cline{8-9} \cline{11-12}
		$ t-t_{0} $&$\alpha_{1}$& $\beta_{1}$&&$\alpha_{2}$&$\beta_{2}$&&$\alpha_{3}$&$\beta_{3}$&&$ \lambda $ & $ \rho $\\
		\hline 
		$ 7 $ &$ -953.3(4.0) $ & $ 106.4(2.2) $ && $ -51.8(2.7) $ & $3.4(4) $ && $-12.2(4.1)$ & $0.635(230)$&& $-31.0(2.0)$ & $36.5(2.5)$   \\
		$ 8 $ &$ -990.3(6.5) $ & $ 106.6(3.2) $ && $ -52.9(2.6) $ & $3.9(6) $ && $-15.2(5.0)$ & $0.791(251)$&& $-32.9(2.8)$ & $36.5(3.4)$   \\
		$ 9 $ &$ -1011.6(8.6) $ & $ 109.8(3.0) $ && $ -57.4(2.4) $ & $4.0(4) $ && $-13.2(4.2)$ & $0.766(253)$&& $-32.8(2.1)$ & $39.2(3.0)$   \\
		\hline
	\end{tabular}
	%\end{ruledtabular}
	{\label{tab:V_LK-LK}}
\end{table}
%%%%%%%%%%%%%%%%%%%%%%%%
\begin{table}
	\caption{
		Fitted parameters in Eq.~\eqref{eq:V_D} for $ V_{pD_{s}}^{pD_{s}} $ with the statistical errors using the data at $ r<1.7 $ fm. Units are the same as those in Table~\ref{tab:V_LK-LK}.
	%	\textcolor{red}{The values of $ \chi^{2}/{d.o.f} $ with $ d.o.f=214 $ are $ 1.30(40),0.76(18) $ and $ 0.74(30) $ for  $ t-t_{0}=7,8 $ and $ 9 $, respectively.}	
	}
	\centering
	%	\begin{ruledtabular}
	\begin{tabular}{ccccccccccccc}
	\hline
	&	  \multicolumn{2}{c}{Gauss-1}    && \multicolumn{2}{c}{Gauss-2}    &&  \multicolumn{2}{c}{Gauss-3} &&  \multicolumn{2}{c}{$ {\textrm{Yukawa}}^{2} $}\\ \cline{2-3} \cline{5-6} \cline{8-9} \cline{11-12}
	$ t-t_{0} $&$\alpha_{1}$& $\beta_{1}$&&$\alpha_{2}$&$\beta_{2}$&&$\alpha_{3}$&$\beta_{3}$&&$ \lambda $ & $ \rho $\\
	\hline 
$ 7 $ &$ -604.5(4.2) $ & $ 98.0(1.6) $ && $ -74.4(2.2) $ & $3.9(4) $ && $-13.4(5.6)$ & $0.848(340)$&& $-24.8(4.4)$ & $25.0(2.9)$   \\
$ 8 $ &$ -621.5(3.9) $ & $ 98.3(1.5) $ && $ -74.8(2.6) $ & $4.1(4) $ && $-14.5(5.8)$ & $0.947(350)$&& $-25.2(3.9)$ & $25.9(2.7)$   \\
$ 9 $ &$ -630.2(3.6) $ & $ 103.5(1.8) $ && $ -78.6(3.5) $ & $4.6(4) $ && $-17.9(5.7)$ & $1.1(3)$&& $-19.0(2.0)$ & $33.9(3.0)$   \\

	\hline
\end{tabular}
	%\end{ruledtabular}
	{\label{tab:V_pDs-pDs}}
\end{table}
%%%%%%%%%%%%%%%%%%%%%%%%
\begin{table}
	\caption{
		Fitted parameters in Eq.~\eqref{eq:vbar} for Hermitian
		$\bar{V}_{\Lambda_{c}K^{+}}^{pD_{s}} $ potential with the statistical errors using the data at $ r<1.7 $ fm. Units are the same as those in Table~\ref{tab:V_LK-LK}.
		}
	\centering
	%	\begin{ruledtabular}
	\begin{tabular}{ccccccccc}
		\hline
		&	  \multicolumn{2}{c}{Gauss-1} && \multicolumn{2}{c}{Gauss-2} &&  \multicolumn{2}{c}{Gauss-3}\\ \cline{2-3} \cline{5-6} \cline{8-9} 
		$ t-t_{0} $&$\alpha_{1}$& $\beta_{1}$&&$\alpha_{2}$&$\beta_{2}$&&$\alpha_{3}$&$\beta_{3}$\\
		\hline 
		$ 7 $ &$ 72.8(1.5) $ & $ 52.9(1.9) $ && $ 23.2(1.5) $ & $8.0(5) $ && $2.8(4)$ & $0.629(133)$\\
		$ 8 $ &$ 71.8(4.7) $ & $ 45.9(4.3) $ && $ 23.6(2.1) $ & $7.8(9) $ && $2.8(9)$ & $0.959(300)$\\
		$ 9 $ &$ 83.1(3.4) $ & $ 55.6(4.8) $ && $ 27.2(2.4) $ & $8.7(1.3) $ && $4.1(1.6)$ & $1.4(4)$\\
		\hline
	\end{tabular}
	%\end{ruledtabular}
	{\label{tab:vbar}}
\end{table}
%%%%%%%%%%%%%%%%%%%%%%%%
\section{Phase shift and scattering length\label{subsec:Phase-shift-and}}
The physical observables such as coupled-channel scattering phase shifts of $ \Lambda_{c}K^{+}-pD_{s} $ interactions can be calculated.  
Therefore, we solve the coupled-channel Schr{\"{o}}dinger equation with the fitted potentials (given in the previous section) in the
infinite volume and extract S- matrix from the asymptotic behaviour
of the wave functions. Here the convention introduced by Stapp et al.~\cite{StapPR1957} is used for the definition of phase shifts in the case of coupled channel system as
\begin{equation}
\left(\begin{array}{cc}
S_{\Lambda_{c}K^{+}}^{\Lambda_{c}K^{+}} & S_{\Lambda_{c}K^{+}}^{pD_{s}}\\
S_{pD_{s}}^{\Lambda_{c}K^{+}} & S_{pD_{s}}^{pD_{s}}
\end{array}\right)=
\end{equation}
$$
\left(\begin{array}{cc}
e^{i\bar{\delta}_{\Lambda_{c}K^{+}}} & 0\\
0 & e^{i\bar{\delta}_{pD_{s}}}
\end{array}\right)\left(\begin{array}{cc}
\cos2\bar{\epsilon} & i\sin2\bar{\epsilon}\\
i\sin2\bar{\epsilon} & \cos2\bar{\epsilon}
\end{array}\right)\left(\begin{array}{cc}
e^{i\bar{\delta}_{\Lambda_{c}K^{+}}} & 0\\
0 & e^{i\bar{\delta}_{pD_{s}}}
\end{array}\right),
$$
where $\bar{\delta}$ is so-called the bar phase shift and $\bar{\epsilon}$
is the mixing angle.

\begin{figure}
	\begin{center}
		\includegraphics[scale=0.7]{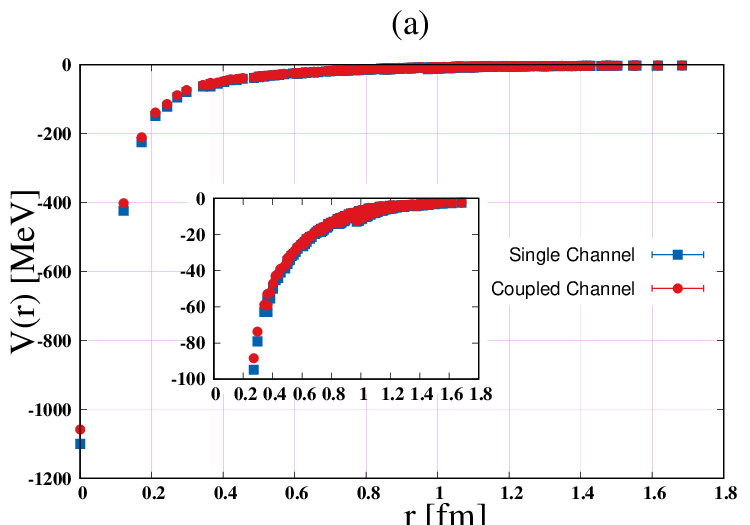}\includegraphics[scale=0.7]{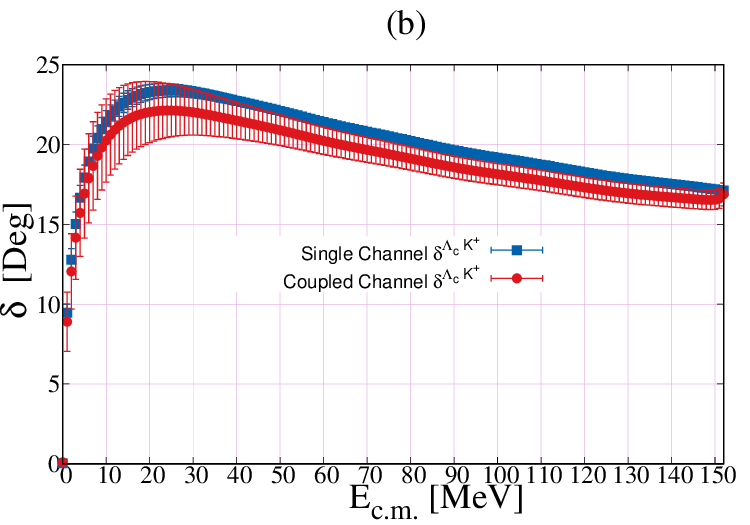} 
		\caption{Panel (a) shows a comparison between 
			the  single channel $ {\Lambda_{c}K^{+}} $ potential (blue square) and a diagonal part of the coupled channel potential (red circle), 
			and panel (b) presents the relevant phase shifts in the single and coupled channel at $ t-t_{0}= 8 $. 
				The mild difference between single and coupled channel phase shift is due to the effect of coupled channel.
		The $\Lambda_{c}K^{+}$  phase shift
			are plotted against the energies from the $\Lambda_{c}K^{+}$ 
			threshold. In the case of single channel $ \Lambda_{c}K^{+} $ potential the corresponding fit parameters are given by Table.~\ref{tab:V_LK_single} in Appendix~\ref{subsec:fit-para-single-channel}. \label{fig:singleVScoupled} }
	\end{center}
\end{figure}

At first, since it is desired to see how large the coupled channel
effect, 
we have compared the single channel $ {\Lambda_{c}K^{+}} $ potential and a diagonal part of the coupled channel potential, i.e., $V_{\Lambda_{c}K^{+}}^{\Lambda_{c}K^{+}}$ at $ t-t_{0}=8 $ (panel (a) in Fig~\ref{fig:singleVScoupled}) and their corresponding scattering phase shifts (panel (b) in Fig~\ref{fig:singleVScoupled}) below $ p D_{s} $ threshold $ (W_{CN} \simeq 157 \textrm{MeV}) $ as well. 
By using the single 
channel potential and solving the Schr{\"{o}}dinger equation,  it is established that the single channel potential does not have a bound state.

\begin{figure}
	\begin{center}
		\includegraphics[scale=0.7]{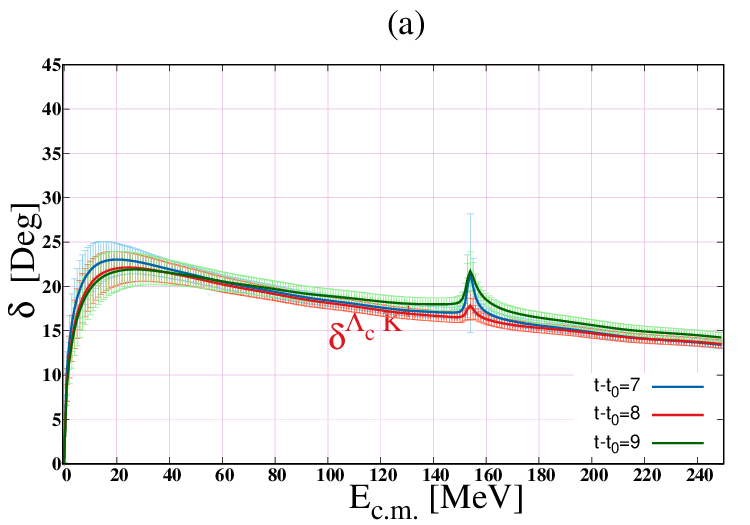}\includegraphics[clip,scale=0.7]{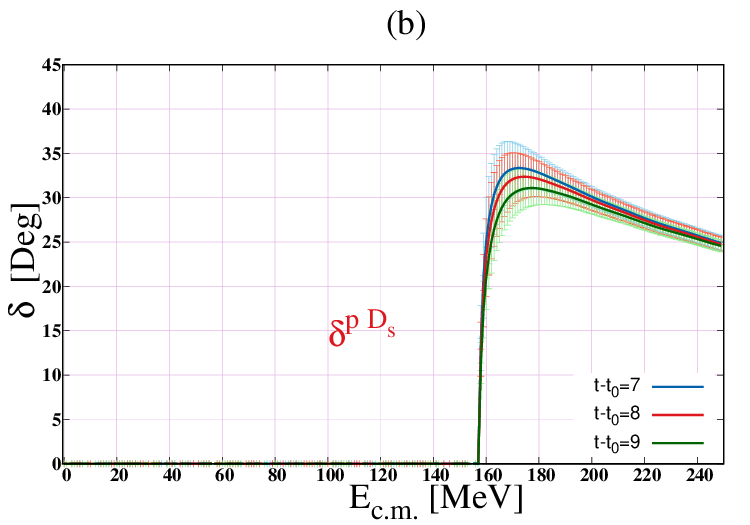}
		
		\includegraphics[clip,scale=0.7]{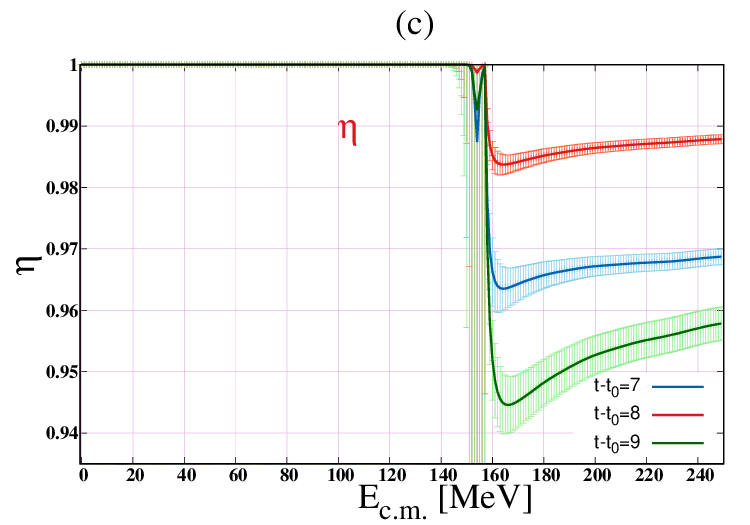}
		\caption{The phase shifts of S-wave (a) $\Lambda_{c}K^{+}$, (b) $pD_{s}$ and 
			%in the\textcolor{red}{{} ${\color{black}I\left(J^{p}\right)=1/2\left(1/2^{-}\right)}$} state. 
			 (c) the inelasticity of the scattering defined as
			  $ \eta =\left|S_{\Lambda_{c}K^{+}}^{\Lambda_{c}K^{+}}\right|=\left|\cos2\bar{\epsilon}\right|$
			is plotted against the energies from the $\Lambda_{c}K^{+}$ threshold for $ t-t_{0}= 7 $ (blue), $ 8 $ (red) and $ 9 $ (green).
			\label{fig:ps} }
	\end{center}
\end{figure}

Panel (a) and panel (b) in Fig. \ref{fig:ps}, shows resultant S-wave scattering
phase shifts for $ \Lambda_{c}K^{+} $  and $ pD_{s} $ channels  as a function of the baryon-meson center-of-mass
energy, respectively. 
The $ \Lambda_{c} K $ phase-shift shows a weak attraction. Moreover, just above the $ \Lambda_{c} K $ threshold the $ S $-wave phase shift shows a rapid variation, a feature indicative of a nearby pole. 
However, at the opening of the $ pD_{s} $ threshold we do observe a noticeable “kink” in the $ \Lambda_{c} K $ phase
shift suggesting a non-zero coupling between the two channels.
The coupled $ \Lambda_{c} K $, $ pD_{s} $ system, showing an enhancement (steadily increasing) in the $ \Lambda_{c} K $ phase shift.
The non-zero coupling between channels is
further demonstrated in Fig.~\ref{fig:ps}, panel (c), which shows a clear deviation of the inelasticities, $ \eta $ from unity.
The inelasticity indicates
that the transition between the $ \Lambda_{c}K^{+} $ and $ pD_{s} $
is weak and two channels are almost independent each other. This observation
might be understood from the fact that the large mass splitting between
the $ \Lambda_{c}K^{+} $ and $ pD_{s} $.

As mentioned above, because the two channels are
almost independent each other due to the small mixing, they can
be regarded as two single channels, therefore, from the low-energy part of $ \Lambda_{c}K^{+}\left(pD_{s}\right) $ phase shifts in Fig.~\ref{fig:ps} (a) ((b)) the scattering length $ \left(a_{0}\right) $ and the effective range $ \left(r_{\textrm{eff}}\right) $ are derived by using the S-wave effective range expansion (ERE) up to the next-to-leading order (NLO),
\begin{equation}
	k\:\cot\delta=-\frac{1}{a_{0}}+\frac{1}{2}r_{\textrm{eff}}k^{2}+\mathcal{O}\left(k^{4}\right),
\end{equation} 
The results are
% scatt_length=-0.890618  +/-  0.210842  +/-  -0.14532  +/-  0.0592134
%Effective-Range=2.63711  +/-  0.29001  +/-  -0.255907  +/-  -0.0282762
\begin{eqnarray}
\begin{array}{cc}
a_{0}^{\Lambda_{c}K^{+}}= -0.87\pm0.19_{-0.12}^{+0.06} \: \textrm{fm}, & r_{\textrm{eff}}^{\Lambda_{c}K^{+}}=2.60\pm0.26_{-0.22}^{+0.03}  \:\textrm{fm},\end{array} \label{eq:ere-Lamc-Kp}
\end{eqnarray}
%scatt_length=-1.42762  +/-  0.372053  +/-  0.173766  +/-  -0.183369
%Effective-Range=1.76406  +/-  0.186747  +/-  -0.0552941  +/-  0.0719479
\begin{eqnarray}
	\begin{array}{cc}
		a_{0}^{pD_{S}}=-1.42\pm0.36_{-0.17}^{+0.18} \: \textrm{fm}, & r_{\textrm{eff}}^{pD_{S}}=1.74\pm0.19_{-0.06}^{+0.07}  \:\textrm{fm},\end{array} \label{eq:ere-pDs}
\end{eqnarray}
%\textcolor{red}{(10) p.15	I do not understand (19), (20) and (21).  More explanations are necessary.}
where the central values and the statistical errors are evaluated at $  t-t_{0}=8 $, while the systematic errors are evaluated by the difference between the central values at $ t-t_{0} $ and those at $ t-t_{0}=7 $ and $ 9 $.
%$ a_{\Lambda_{c}K^{+}}=1.365(935)+ i 0.014(11) $ and $ a_{pD_{s}}= 0.039(16)+i 0.006(3) $. Here the errors are statistical errors, and for the scattering phase shifts are calculated by the jackknife method. 

In addition, because the electric charge of each hadron in the  $\Lambda_{c}K^{+}-pD_{s}$ system is $ +e $ we have Coulomb repulsion. 
It is probably too early to consider Coulomb interaction since the pion mass in our simulation is as heavy as $ 872 $ MeV, 
%
%but in order to have an estimation, if we switch on the Coulomb interaction by adding the correction $ V_{\textrm{Coul}}\left(r\right)=\alpha_{f}/r $ to QCD potential matrix elements, where $ \alpha_{f}=1/137.036 $ is the fine structure constant, the results with coulomb potential, $ a_{0,C} $ and $ r_{\textrm{eff,C}} $, become
%scatt_length=1.13417  +/-  0.0405551  +/-  0.0248479  +/-  -0.0102382
%Effective-Range=24.2837  +/-  6.87535  +/-  -0.36116  +/-  1.23188
%\begin{eqnarray}
%	\begin{array}{cc}
%		a_{0,C}^{\Lambda_{c}K^{+}}= 1.13\pm0.04_{+0.02}^{-0.01} \: \textrm{fm}, & r_{\textrm{eff,C}}^{\Lambda_{c}K^{+}}=24\pm7_{-1.0}^{+1.4}  \:\textrm{fm},\end{array} \label{eq:ere-Lamc-Kp}
%\end{eqnarray}
%scatt_length=1.5836  +/-  0.0761035  +/-  0.00106815  +/-  -0.0135261
%Effective-Range=13.3234  +/-  2.48225  +/-  0.551648  +/-  -0.463145
%\begin{eqnarray}
%	\begin{array}{cc}
%		a_{0,C}^{pD_{S}}=2.07\pm0.06_{+0.03}^{-0.03} \: \textrm{fm}, &
%		 r_{\textrm{eff,C}}^{pD_{S}}=-21\pm5_{+2}^{-2}  \:\textrm{fm},\end{array} \label{eq:ere-pDs}
%\end{eqnarray}
therefor for future lattice QCD simulation at physical point (with nearly physical quark mass, i.e., $ m_{\pi}\simeq146 $ MeV and $ m_{K}\simeq525 \textrm{MeV} $)~\cite{Ishikawa:2015rho},
 it is necessary to include Coulomb interaction.  
 %~\cite{Ukita:2016d4},

\section{Summary and conclusion\label{sec:Summary-and-conclusion}}
We have investigated the S-wave $ \Lambda_{c}K^{+} $ interaction 
%in the \textcolor{red}{${\color{black}I\left(J^{p}\right)=1/2\left(1/2^{-}\right)}$} state
 using the $ \Lambda_{c}K^{+}-pD_{s} $ coupled channel potentials
obtained by the extension of the HAL QCD method in Flavor $ \textrm{SU}\left(3\right) $ Limit of Lattice QCD. 
Results of the potential matrix show that the diagonal elements, $\Lambda_{c}K^{+}$ and $pD_{s}$ 
are both strongly attractive, while, $\Lambda_{c}K^{+}$ has a much deeper attractive pocket than $pD_{s}$.
We have also observed
weak off-diagonal elements of the potential matrix. They are not
Hermitician, which is not guaranteed
from the definition. The phase shifts and inelasticity extracted by
solving Schr{\"{o}}dinger equation in the infinite volume with the obtained
potentials show that the $\Lambda_{c}K^{+}$ channel does not have
the two-body bound state at $ m_{\pi}\geq872 $ MeV. While the inelasticity (mixing angle
between $\Lambda_{c}K^{+}$ and $pD_{s}$ channel also indicates a
small coupling between the two.
the coupled $ \Lambda_{c} K $, $ p D_{s} $ system, showing an enhancement (steadily increasing) in the $ \Lambda_{c} K $ phase shift.

And last but not the least,  since in one hand the $ pD_{s}^* $ threshold is quite close to the
kinematic region considered, and on the other hand, the mass difference between  $ D_{s} $ ($ 1968.30 $ MeV) and $ D_{s}^*$ ($ 2112.1 $ MeV)  is about $ 143.8  $ MeV, therefore, in future it is necessary to carry out full channel coupling analysis with $ pD_{s}^* $ by using  $ \left(2+1\right) $-flavor and $ \left(1+1+1\right) $-flavor lattice simulation at physically quark masses  $ \left(m_{\pi}\approx138\:\textrm{MeV}\right)  $ on a sufficient large lattice.

% In future, we will include the $ pD_{s}^{\ast} $ ($D_{s}^{*}$($c\bar{s})$)  into our calculation.
%We will also calculate the interaction of the $\Lambda_{c}K^{+}$ system at physical pion mass.

\section*{Acknowledgement}
We thank the members of HAL
QCD Collaboration for technical supports and helpful discussions.
We thank CP-PACS/JLQCD Collaborations and ILDG/JLDG ~\cite{JLDG} for providing us the $ \textrm{SU}\left(3\right) $-flavor
gauge configurations.
F.E.  thanks Yukawa Institute for Theoretical Physics, Kyoto University for a kind hospitality
during his stay while completing calculations.

\appendix
\section{Dependence on different off-diagonal potentials}\label{subsec:more_analysis}

 Since the  Hermiticity is broken at short distances, the low energy scattering observables does not suffer from difference between $ V_{\Lambda_{c}K^{+}}^{pD_{s}} $ and $ V_{pD_{s}}^{\Lambda_{c}K^{+} } $. We have checked this by performing the following analysis. 
We first made the coupled channel potential Hermitian (an average or take one of them for the off-diagonal part).
%$ V_{pD_{s}}^{\Lambda_{c}K^{+} } $ and $ V_{\Lambda_{c}K^{+}}^{pD_{s}} $
% We then performed the analysis with coupled channel potential even above the threshold.
We then performed the analysis with coupled channel potential even above the threshold to check the observable difference among all 4 following cases of the off-diagonal part (one non-Hermitian and three Hermitian)

\begin{figure}
	\begin{center}
		\includegraphics[scale=0.7]{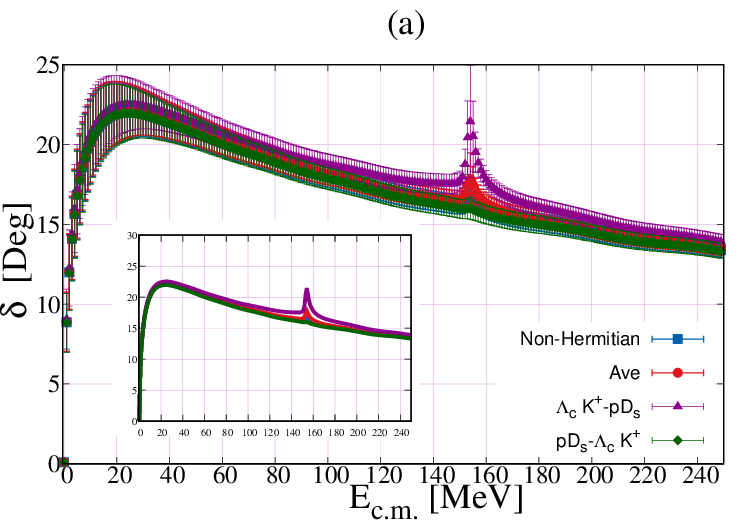}\includegraphics[clip,scale=0.7]{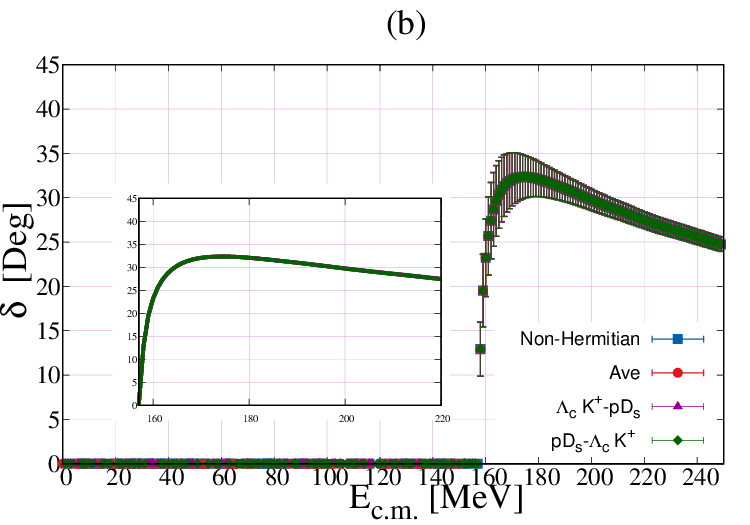}		
		\includegraphics[clip,scale=0.7]{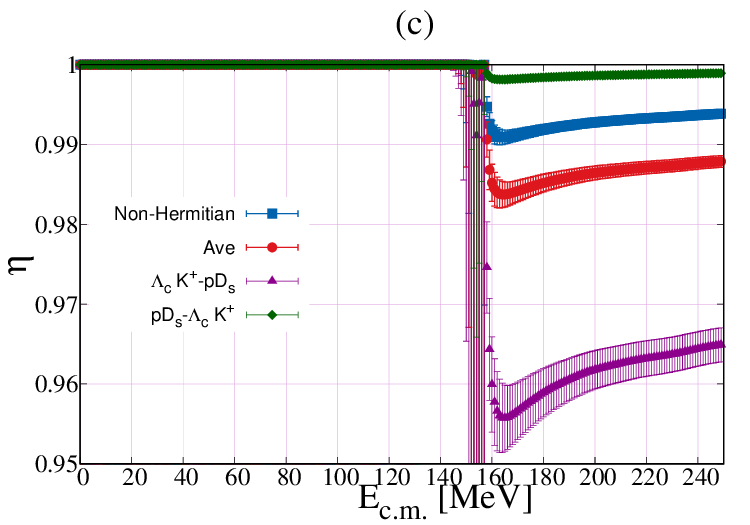}
		\caption{The scattering phase shifts of S-wave (a) $\Lambda_{c}K^{+}$, (b) $ pD_{s} $  and (c) the inelasticity parameter correspond to the phase shifts calculated with non-Hermitian (blue), $ \bar{V}_{\Lambda_{c}K^{+}}^{pD_{s}} $ (red),
		$  V_{\Lambda_{c}K^{+}}^{pD_{s}}  $ (purple) and
		 $ V_{pD_{s}}^{\Lambda_{c}K^{+} } $ (green) at $ t-t_{0}=8 $. 		
		% of the scattering defined as $\left|S_{\Lambda_{c}K^{+}}^{\Lambda_{c}K^{+}}\right|=\left|\cos2\bar{\epsilon}\right|$
		%	is plotted against the energies from the $\Lambda_{c}K^{+}$ threshold.
		\label{fig:ps-abcd} }
	\end{center}
\end{figure}

(a)  Non-Hermitian,
(b) Average of off-diagonal part  i.e., $ \bar{V}_{\Lambda_{c}K^{+}}^{pD_{s}} $,
(c) $V_{12}=V_{21}=V_{\Lambda_{c}K^{+}}^{pD_{s}} $,
(d) $ V_{12} = V_{21} =V_{pD_{s}}^{\Lambda_{c}K^{+} }  $.
The scattering phase shifts of $ {\Lambda_{c}K} $ and $ {pD_{s}} $ and the relevant inelasticity parameter for the above $ 4 $ cases are presented and compared in panel (a), (b) and (c) of Fig.~\ref{fig:ps-abcd}, respectively.			

%%%%%%%%%%%
\section{Fitted parameters for single channel $ \Lambda_{c}K^{+} $ potential}\label{subsec:fit-para-single-channel}
The analytic form same as Eq.~\eqref{eq:V_D} also is used for single channel $ \Lambda_{c}K^{+} $ potential. The resultant fit parameters are given in Table.~\ref{tab:V_LK_single}. The scattering phase shift which obtained by these parameters for $ t-t_{0}=8 $, is presented in Fig.~\ref{fig:singleVScoupled}, panel (b).
	%%%%%%%%%%%%%%%%%%%%%%%%
	\begin{table}
		\caption{
			Fitted parameters in Eq.~\eqref{eq:V_D} for single channel $ \Lambda_{c}K^{+} $ potential with statistical errors using the data at $ r<1.7 $ fm.
			$ \alpha_{i},\beta_{i} $ and $ \rho $ are given in units of 
			$ \left[\textrm{MeV}\right],\left[\textrm{fm}^{-2}\right] $ and
			$ \left[\textrm{fm}^{-2}\right] $, respectively. 
			%\textcolor{red}{The values of 	$ \chi^{2}/{d.o.f} $ with $ d.o.f=214 $ are $ 1.30(40),0.76(18) $ and $ 0.74(30) $ for $ t-t_{0}=7,8 $ and $ 9 $, respectively.}
		}
		\centering
		%	\begin{ruledtabular}
			\begin{tabular}{ccccccccccccc}
				\hline
				&	  \multicolumn{2}{c}{Gauss-1}    && \multicolumn{2}{c}{Gauss-2}    &&  \multicolumn{2}{c}{Gauss-3} &&  \multicolumn{2}{c}{$ {\textrm{Yukawa}}^{2} $}\\ \cline{2-3} \cline{5-6} \cline{8-9} \cline{11-12}
				$ t-t_{0} $&$\alpha_{1}$& $\beta_{1}$&&$\alpha_{2}$&$\beta_{2}$&&$\alpha_{3}$&$\beta_{3}$&&$ \lambda $ & $ \rho $\\
				\hline 
				$ 7 $ &$ -632.7(3.0) $ & $ 102.5(2.0) $  && $-44.6(1.2)$ & $3.42(27)$&& $ -14.5(2.3) $ & $0.47(10) $ &&  $-24.6(1.8)$ & $31.9(2.2)$   \\
				$ 8 $ &$ -1027.5(7.5) $ & $ 106.0(3.0) $ && $-56.6(2.3)$ & $4.02(51)$&&$ -16.1(4.8) $ & $0.80(23) $ && $-32.9(2.8)$ & $36.5(3.4)$   \\
				$ 9 $ &$ -1048.6(8.8) $ & $ 108.4(4.2) $ && $-61.2(2.2)$ & $4.01(411)$&&$ -13.9(4.2) $ & $0.76(0.24) $ &&  $-36.2(2.5)$ & $37.6(3.0)$   \\
				\hline
			\end{tabular}
			%\end{ruledtabular}
			{\label{tab:V_LK_single}}
		\end{table}
		%%%%%%%%%%%%%%%%%%%%%%%%
\bibliography{Refs}% Produces the bibliography via BibTeX.
\end{document}